\documentclass[aps,prl,reprint]{revtex4-1}

\usepackage[mathlines]{lineno}
\usepackage{lipsum}
\usepackage{blindtext}
\usepackage[export]{adjustbox}
\usepackage{graphicx,epsfig,epstopdf}
\usepackage{amsmath,amssymb}
\usepackage{esvect}
\usepackage{mathtools}
\usepackage{color}
\usepackage{subfigure}
\usepackage{array}
\usepackage{tabularx}
\usepackage{multirow}
\usepackage{booktabs}
\usepackage{float}
\newcolumntype{L}[1]{>{\raggedright\arraybackslash}p{#1}}
\newcolumntype{C}[1]{>{\centering\arraybackslash}p{#1}}
\newcolumntype{M}[1]{>{\centering\arraybackslash}m{#1}}
\allowdisplaybreaks

\begin{document}

\title{Extreme Asymmetry in Metasurfaces via Evanescent Fields Engineering: Angular-Asymmetric Absorption}

\author{Xuchen Wang}
\author{Ana~D\'{i}az-Rubio}
\author{Viktar S. Asadchy}
\author{Grigorii Ptitcyn}
\author{Andrey A. Generalov}
\author{Juha Ala-Laurinaho}
\author{Sergei A. Tretyakov}

\affiliation{Department of Electronics and Nanoengineering, Aalto University, Finland}

\date{\today}

\begin{abstract}
On the quest towards full control over wave propagation, the development of compact devices that allow asymmetric response is a challenge. In this Letter, we introduce a new paradigm for the engineering of asymmetry in planar structures, revealing and exploiting unilateral excitation of evanescent waves. We test the idea with the design and experimental characterization of a metasurface for angular-asymmetric absorption. The results show that the contrast ratio of absorption (the asymmetry level) can be arbitrarily engineered from zero to infinity for waves coming from two oppositely tilted angles. We demonstrate that the revealed asymmetry effects  cannot be realized using conventional diffraction gratings,  reflectarrays, and phase-gradient metasurfaces. This Letter opens up promising possibilities for wave manipulation via evanescent waves engineering with applications in one-side detection and sensing, angle-encoded steganography, flat nonlinear devices and shaping the scattering patterns of various objects. 
\end{abstract}

\maketitle

	Asymmetric wave propagation has attracted significant interest in recent years due to its fundamental importance in many applications, for example, in  one-way communication systems. Many asymmetric properties (Faraday rotation, wave isolation, etc.)  can be realized only using nonreciprocal components (e.g., ferrites~\cite{adam2002ferrite}, nonlinear materials \cite{fallahi2012manipulation,mahmoud2015all}, and materials with space-time modulated  properties~\cite{sounas2013giant,hadad2015space}). However, some asymmetric propagation effects (asymmetric wave transmission, reflection, absorption) are \textit{reciprocal} effects. Some interesting results have been achieved using artificial 2D composites, so-called metasurfaces~\cite{menzel2010asymmetric,mutlu_diodelike_2012,
huang_asymmetric_2012,zhang_interference_2013,pfeiffer_bianisotropic_2014,fedotov_asymmetric_2006,wu_giant_2013,pfeiffer2014high,ra2014tailoring,alaee2015all,asadchy_functional_2015,shevchenko_bifacial_2015,odit_experimental_2016,yazdi2015bianisotropic,alaee2015magnetoelectric,faniayeu_vertical_2017}. Depending on the physics of the asymmetric response, the known  structures can be classified to two main types. In the metasurfaces of the first type, asymmetric transmission or reflection is achieved due to asymmetric polarization conversion~\cite{menzel2010asymmetric,mutlu_diodelike_2012,huang_asymmetric_2012,zhang_interference_2013,pfeiffer_bianisotropic_2014,fedotov_asymmetric_2006,wu_giant_2013,pfeiffer2014high}. For example, a right circularly polarized  beam is transmitted from medium~1 to medium~2 as a left circularly polarized beam, while the same right circularly polarized beam,  when incident reciprocally from medium~2 to medium~1,   is fully reflected, preserving its polarization state. Although such system exhibits unidirectional transmission for waves impinging on the two opposite sides, its scattering matrix is symmetric, as dictated by the reciprocity theorem. In metasurfaces of the second type, asymmetric response  can be achieved without polarization conversion,  but only for devices working in the reflection regime. In such devices, the asymmetry of reflection coefficient phase for illuminations from opposite sides was exploited in specifically designed bianisotropic metasurfaces~\cite{ra2014tailoring,alaee2015all,asadchy_functional_2015,shevchenko_bifacial_2015,odit_experimental_2016,asadchy_bianisotropic_2018}. In lossy structures, asymmetric absorption becomes possible~\cite{yazdi2015bianisotropic,alaee2015magnetoelectric,faniayeu_vertical_2017}.

In the aforementioned works, asymmetric wave propagation was achieved only from different sides of composite layers and for normally incident waves. From the theoretical point of view and for many applications (e.g., multifunctional grating, flat optics), it is important to create planar structures with asymmetric response for illuminations from two half-spaces with respect to the normal direction of the surface or for two arbitrary angular sectors.
In principle, if the metasurface allows induced currents in the normal to the surface plane direction, asymmetric reflection at oblique incidence angles can be realized simply by using arrays of tilted elements. An example is an array of small tilted mirrors, which looks transparent if we look along the mirror planes, but strongly reflective if we look from the orthogonal direction. More advanced  angular-asymmetric wave propagation was reported in Ref. \cite{pfeiffer2016emulating}. 
However, this \textit{geometry-induced} mechanism  has significant limitations: the desired performance can be achieved only for $\theta_{\rm i}=45^\circ$ angle, and fabrication of nonplanar structures with tilted inclusions can be complicated. To the best of our knowledge, asymmetrically reflecting planar metasurfaces without  normal polarization are not known.  

In this Letter, we  demonstrate that evanescent waves engineering can provide a robust platform to control the asymmetric response of metasurfaces for oblique incidence, as well as allow other new functionalities.	As an example, we theoretically and experimentally validate angular-asymmetric  absorption in \textit{impenetrable planar reciprocal} surfaces, and we show that the asymmetry level and the incidence configuration can be arbitrarily defined.  This structure exhibits extremely  broken symmetry of absorption properties with respect to the incidence angle: nearly \textit{zero} and nearly \textit{total} absorption for waves incident from $+\theta_{\rm i}$ and $-\theta_{\rm i}$ angles, respectively. 
We demonstrate that, to achieve extreme angular asymmetry of absorption, one must ensure excitation of a proper set of evanescent waves propagating  along the surface in the direction of the incident-wave wave vector.
Moreover, the surface properties  must vary quasicontinuously with discretization of at least 10~elements per one wavelength. Thus, conventional diffraction gratings~\cite{hutley_diffraction_1982,popov_gratingsgeneral_1990} and recently introduced metagratings ~\cite{radi_metagratings_2017,epstein_unveiling_2017,radi_reconfigurable_2018}, which include not more than two~elements per period (three~elements per  wavelength), cannot be exploited to create such asymmetric reflectors.
 Our design of an angular asymmetric absorber is based on gradient metasurfaces~\cite{yu_light_2011,asadchy2017flat}, i.e., diffraction gratings with  properties  varying at a \textit{deeply subwavelength} scale.

\begin{figure}[h!]
	\centering
	\includegraphics[width=0.7\linewidth]{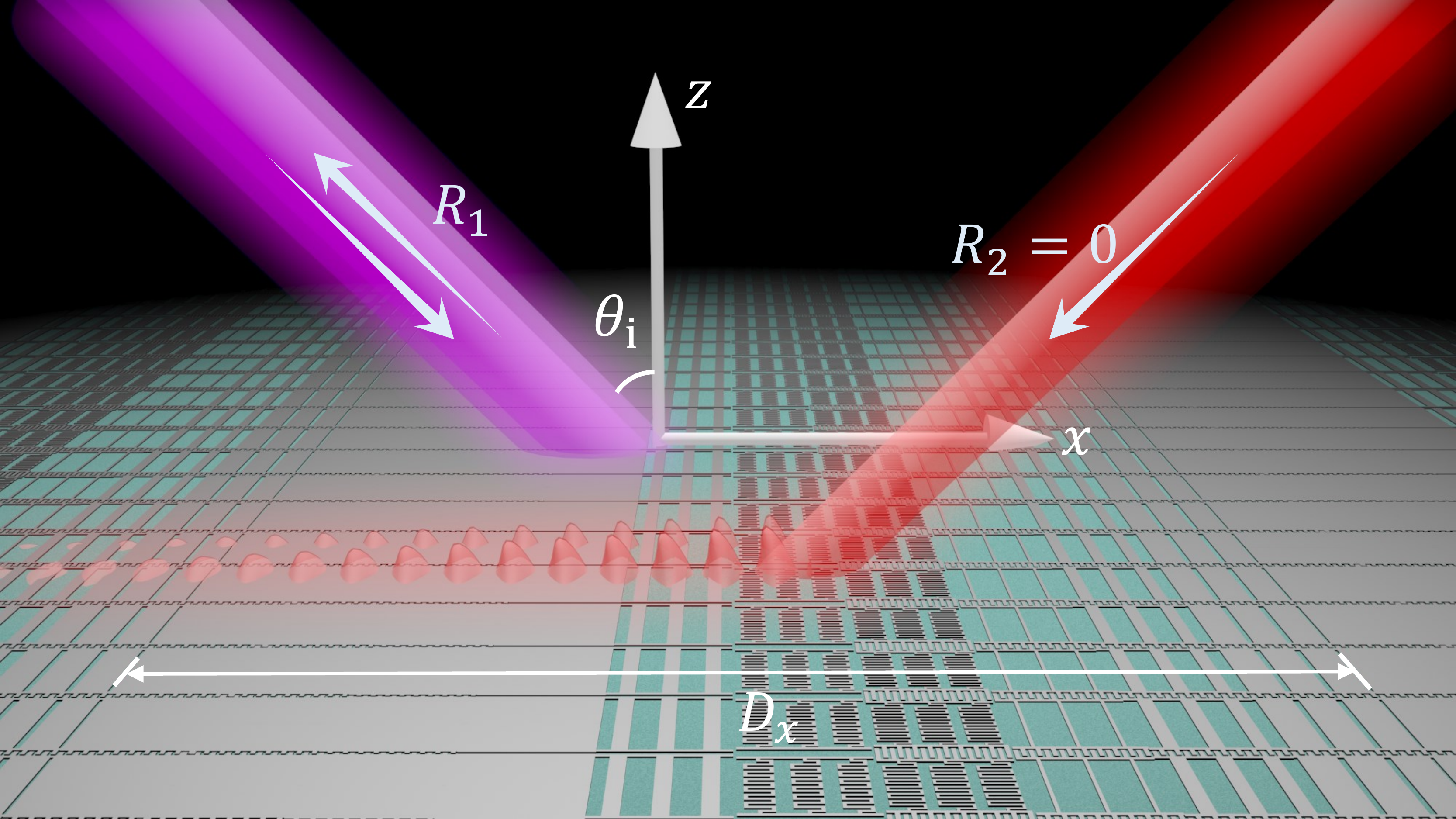}
	\caption{Schematic of the asymmetric absorber. $R_2$ is the reflection coefficient for illumination of $\theta=-\theta_{\rm i}$. }\label{fig:fig1}
\end{figure}
Let us   consider an impenetrable reciprocal  metasurface with periodicity $D_x$ along the $x$ direction, located in the $xy$ plane (see Fig.~\ref{fig:fig1}). 
The metasurface is illuminated by a TE($s$)-polarized  plane wave  at an incidence angle $\theta$. 
Naturally, if the periodicity is smaller than   a half-wavelength of the incident radiation (which is the case for conventional electromagnetic absorbers), the reflection and absorption are even functions with respect to angle~$\theta$. 
This fact originates from reciprocity of the system, which implies that   specular reflections are the same for illuminations from  $+\theta$ and  $-\theta$ angles. 
In order to break the reflection symmetry of the metasurface, it is necessary (but not sufficient) to ensure its proper periodicity so that diffracted modes can propagate. In this case, waves incident from the opposite directions ($+\theta$ and  $-\theta$) can be    reflected into diffracted harmonics and with different amplitudes. Therefore, although the specular reflection remains an even  function of the incidence angle (due to reciprocity), absorption coefficient and \textit{total} reflection become noneven functions. 
The field reflected by a periodically modulated metasurface  can be interpreted as a sum of Floquet harmonics. The tangential wave number of the $n$th harmonic  is related to the period $D_x$ and the incident wave number $k_{\rm 0}$   as $k_{{\rm r}x}=k_{\rm 0}\sin\theta+{2\pi n}/{D_x}$. 
The corresponding normal component of the reflected wave number equals $k_{{\rm r}z}=\sqrt{k_{\rm 0}^2-k_{{\rm r}x}^2}$. If $|k_{{\rm r}x}|$ is greater than the incident wave number, the wave is evanescent and  it does not contribute to the far field. For the harmonic wave  satisfying $|k_{{\rm r}x}|<k_{\rm 0}$, $k_{{\rm r}z}$ is real, and   this wave is propagating.
Our aim is to achieve controllable nonspecular reflection for waves incident on the metasurface at an angle $\theta=+\theta_{\rm i}$  and  full absorption of waves incident at an angle $\theta=-\theta_{\rm i}$. We will show that such asymmetric absorption can be achieved through unbalanced excitation of evanescent waves along the metasurface in these two cases (stronger evanescent waves contribute to higher energy dissipation  at a lossy metasurface). To this end, we look for a solution when a few or none evanescent waves are excited for illumination at $+\theta_{\rm i}$, while  the opposite illumination results in multiple and strong evanescent harmonics.

First, let us consider illumination of the metasurface at an angle $\theta=+\theta_{\rm i}$ and require that the   reflected field is represented only by one propagating harmonic $n=-1$ with electric field amplitude $E_{\rm r}=R_1E_{\rm i}$, where $E_{\rm i}$ is the incident electric field and $R_1$ is the reflection coefficient into that harmonic. Thus, we assume that no specular reflections are allowed and that no evanescent waves are excited for this illumination (in Ref. \cite{suppl}, we demonstrate that this scheme is the most efficient way to realize extremely asymmetric absorption).
The tangential components of the total electric field at the plane $z=0$ can be written as $E_{\rm 1t}(x)=E_{\rm i}(e^{-jk_0\sin\theta_{\rm i}x}+R_1e^{jk_0\sin\theta_{\rm i}x})$, where the
time-harmonic dependency in the form $e^{j\omega t}$ is assumed.
\begin{figure*}[bt!]
	\centering
\subfigure[]{\includegraphics[width=0.3\linewidth]{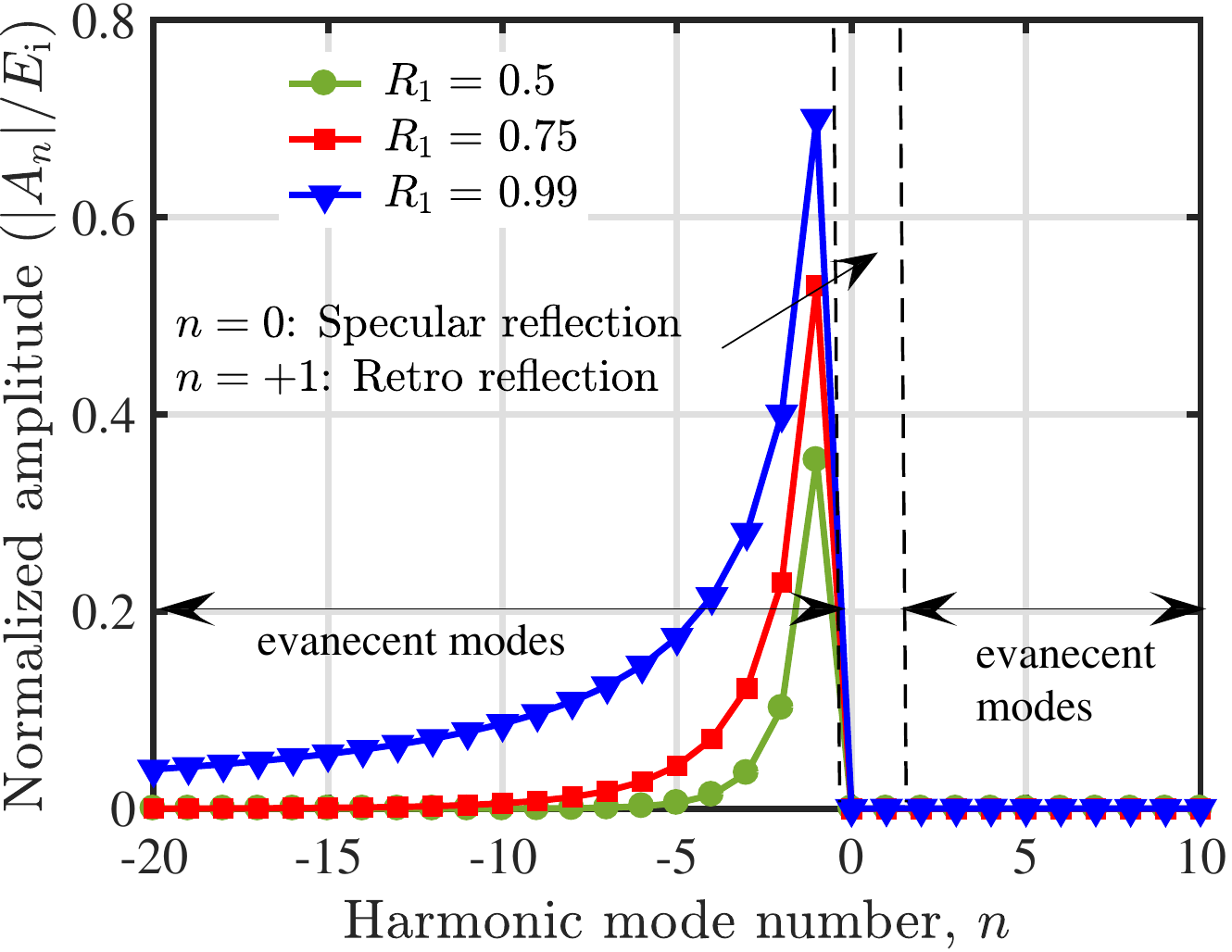}\label{fig:fig2a}}
\subfigure[]{\includegraphics[width=0.3\linewidth]{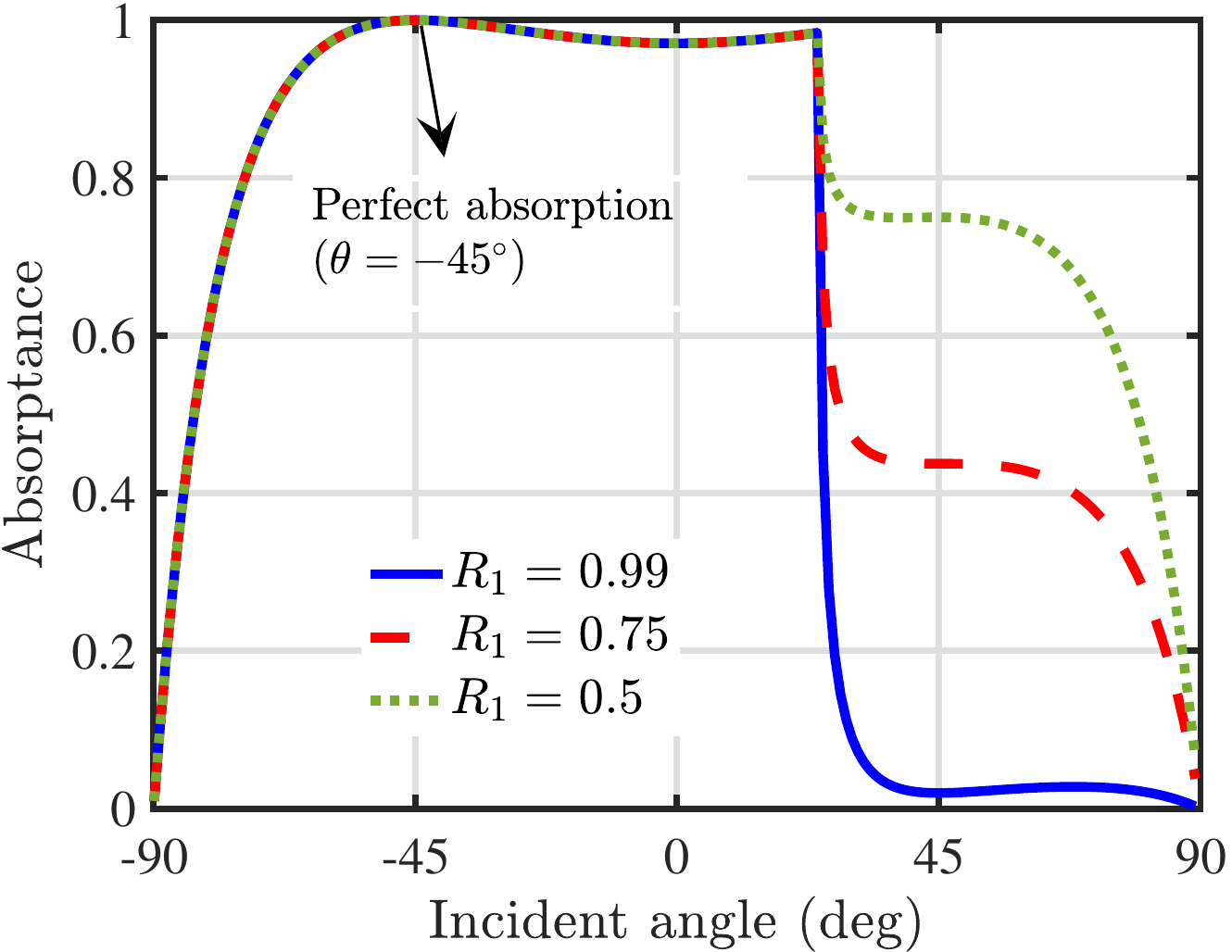}\label{fig:fig2b}}
\subfigure[]{\includegraphics[width=0.3\linewidth]{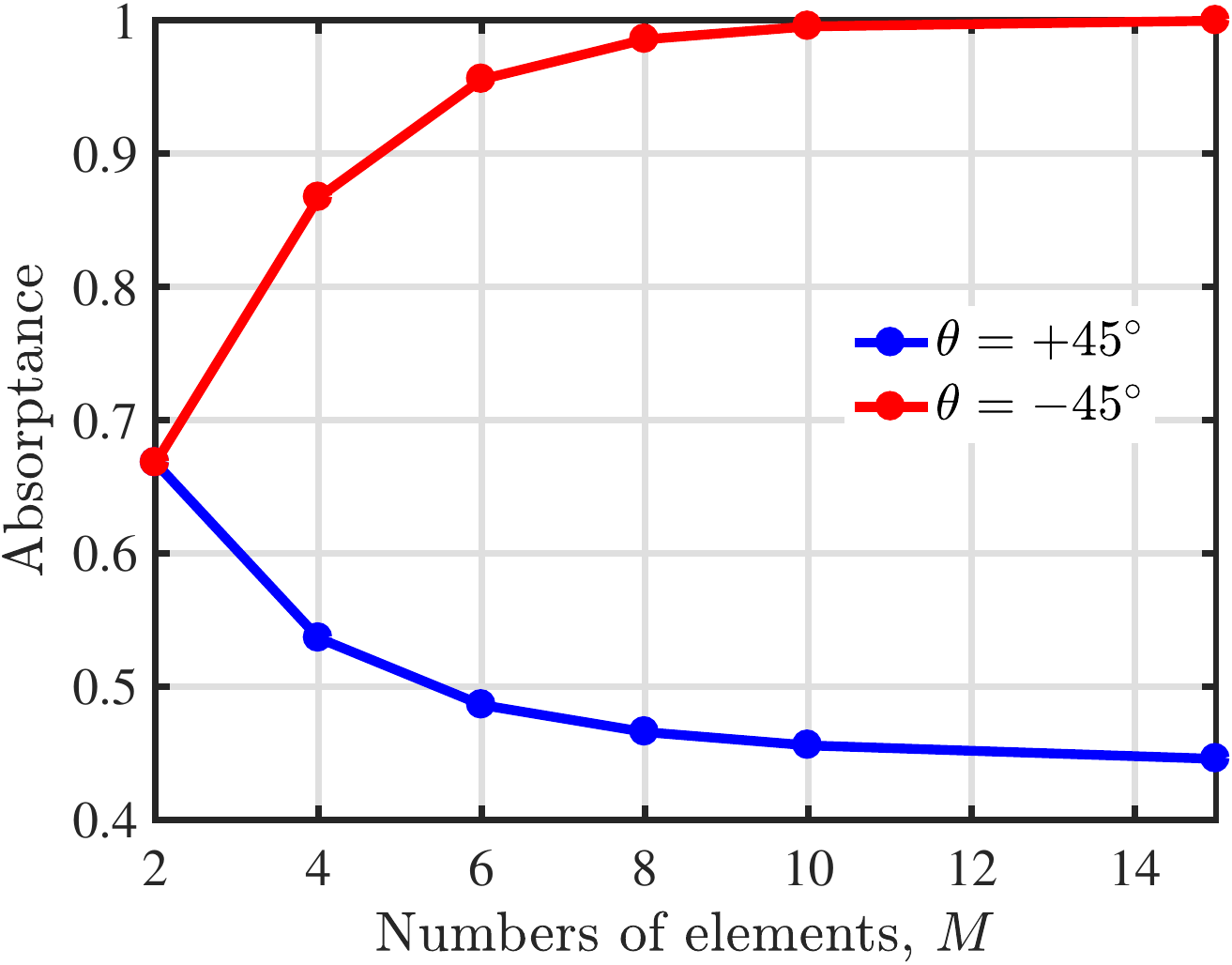}\label{fig:fig2c}}
\subfigure[]{\includegraphics[width=0.28\linewidth]{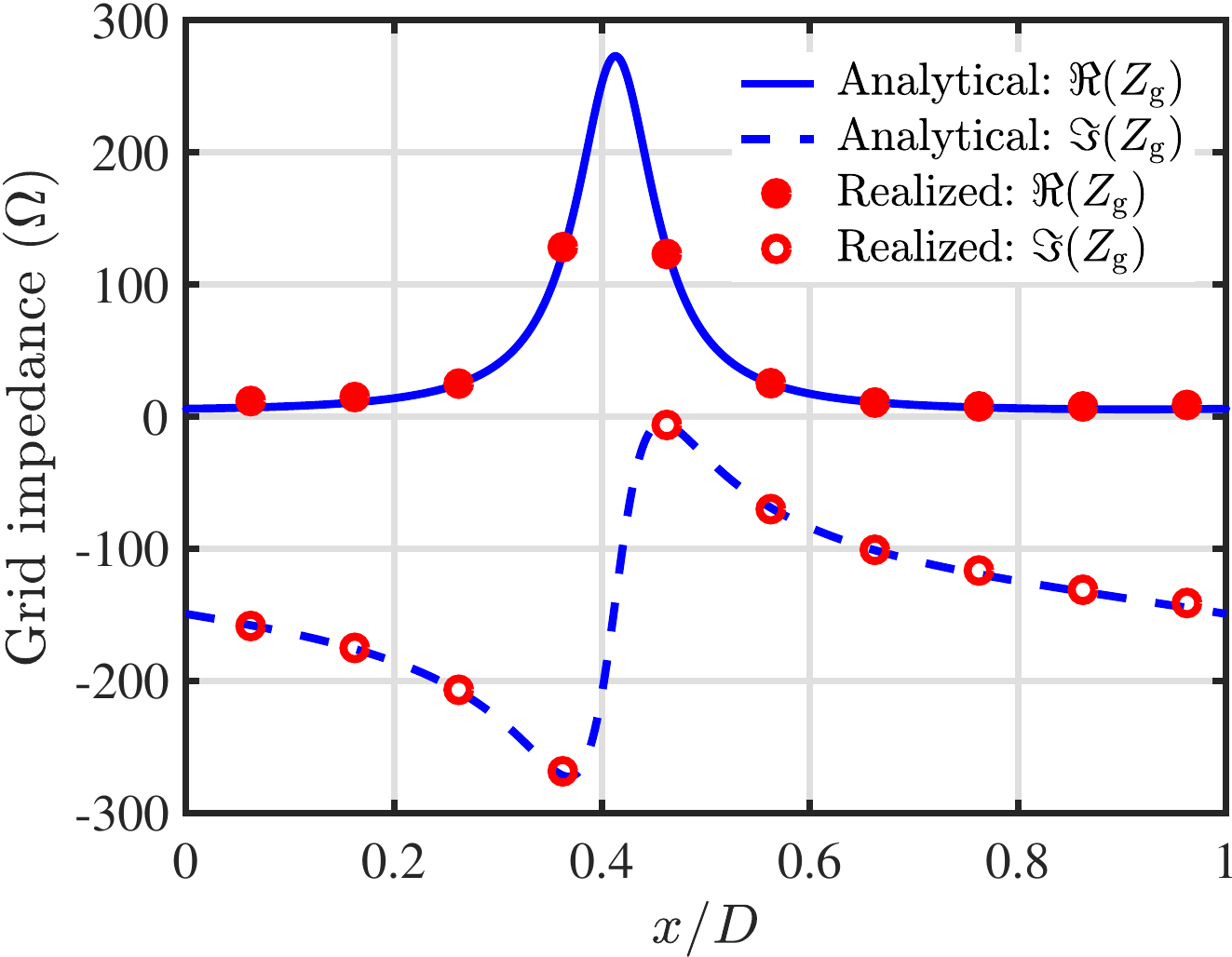}\label{fig:fig2d}}
\subfigure[]{\includegraphics[width=0.21\linewidth]{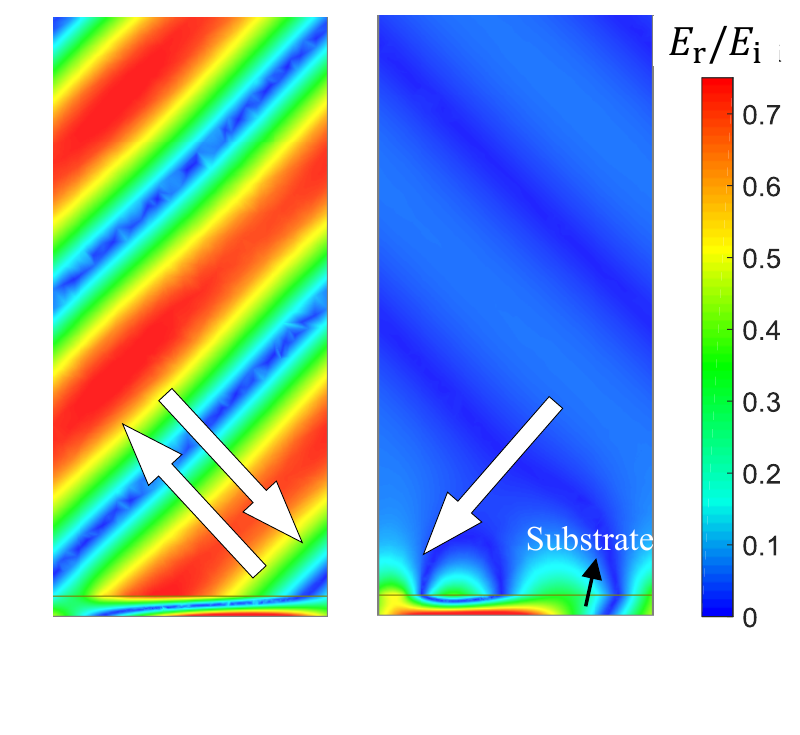}		\label{fig:fig2e}}
\subfigure[]{\includegraphics[width=.43\linewidth]{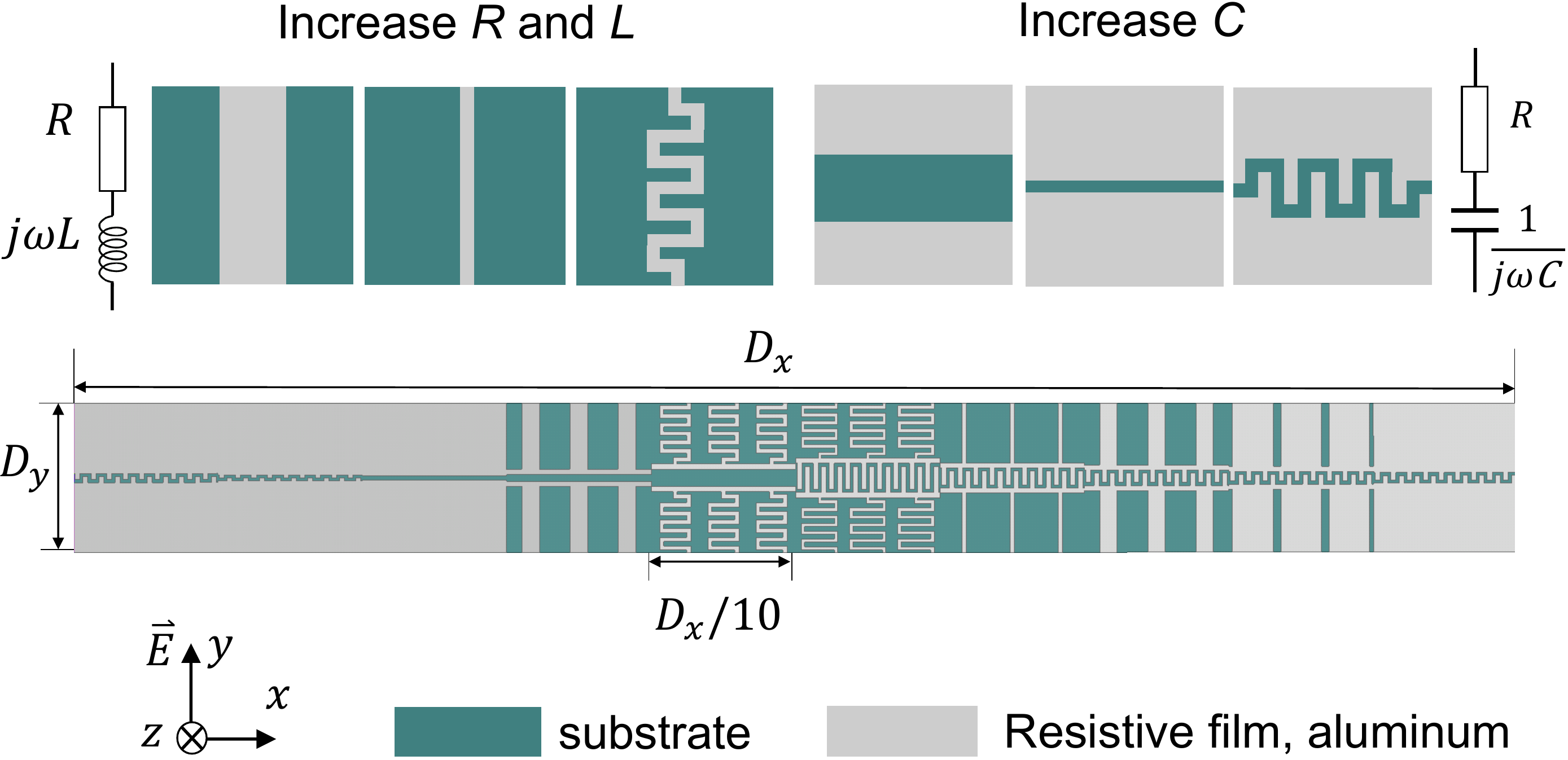}\label{fig:fig2f}}
\caption{ 
(a)    Magnitudes  of the complex amplitudes of the different Floquet harmonics (normalized by the incident electric field) corresponding to three different $R_1$ when the gradient metasurface is illuminated at $\theta=-45^\circ$. (b) Numerically calculated absorptance of the gradient metasurface as a function  of the incident angle for the same three scenarios. (c) Absorptance for illuminations at opposite angles versus the number of elements $M$ over one period (for the case of $R_1=0.75$). (d) Analytical and discretized grid impedance over one period ($R_1=0.75$). (e) Reflected field distribution for the metasurface modeled by the required grid impedance (step-wise approximation, 10 subcells per period) on top of a metal-backed dielectric substrate. Illumination is at  $+45^\circ$ (left) and  at  $-45^\circ$ (right). (f)  Illustration of the method for synthesizing arbitrary grid impedance. Top left: Evolution of the subcell geometry (from left to right) which implies simultaneous increase of the grid resistance $R$ and inductance $L$. Top right: Same for increase of the grid capacitance $C$.
Bottom: unit cell structure of the gradient metasurface consisting of 10~subcells. The unit-cell dimensions are $D_x=2828\ \mu$m  and $D_y=290\ \mu$m.  } 
\end{figure*}
The corresponding total magnetic field reads $H_{\rm 1t}(x)=E_{\rm i}\cos \theta_{\rm i}(e^{-jk_0\sin\theta_{\rm i}x}-R_1e^{jk_0\sin\theta_{\rm i}x})/Z_0$, with $Z_0=\sqrt{\mu_0/\epsilon_0}$ being the free-space wave impedance. 
The ratio of these tangential  electric and magnetic fields gives the required surface impedance  
\begin{equation}
	Z_{\rm s}(x)=\frac{Z_0 (1-R_1^2)/\cos \theta_{\rm i}}{1+R_1^2-2R_1\cos\Phi}+j\frac{2Z_0 R_1\sin\Phi /\cos \theta_{\rm i}}{1+R_1^2-2R_1\cos\Phi}, \label{eq:surface impedance}
\end{equation}
where $\Phi=2k_0\sin\theta_{\rm i}x$ is the linearly varying phase.
The real part of the surface impedance is an even function of $x$     and always non-negative for passive ($|R_1| \le 1$) metasurfaces. In contrast, the imaginary part of the surface impedance  is an odd function  of $x$ which creates   angular asymmetry of the electromagnetic response. As is seen from Eq. (\ref{eq:surface impedance}), the periodicity of the surface impedance is $D_x=\lambda_0/(2\sin\theta_{\rm i})$, where $\lambda_0$ stands for the  wavelength. Arrays with such periodicity   allow only one propagating  diffraction harmonic, and this scenario is sometimes referred in the literature as the Littrow configuration~\cite{loewen_grating_1977,popov_gratingsgeneral_1990,destouches_99_2005} or retroreflection regime~\cite{doumanis_design_2013}. 

If the   metasurface with properties dictated by Eq.~(\ref{eq:surface impedance}) is illuminated at the opposite angle $\theta=-\theta_{\rm i}$ (same polarization), the reflected fields do not contain a single plane wave anymore. Indeed, the incident field has reversed phase distribution $E_{\rm i}e^{+jk_0\sin\theta_{\rm i}x}$, and, therefore,
fulfillment of   the boundary condition~[Eq. (\ref{eq:surface impedance})] is possible only if a certain set of evanescent waves is  excited. These evanescent waves will contribute to additional power dissipation in the metasurface.
The total reflected fields can be represented as an infinite sum of Floquet harmonic modes, $E_{\rm r}=\sum_{n=-\infty}^{\infty}A_ne^{-jk_{{\rm r}z}z}e^{-jk_{{\rm r}x}x}$,
where $A_n$ is the  complex amplitude of the $n$th harmonic. 
The total tangential electric and magnetic fields at  $z=0$  read 
\begin{equation} \begin{array}{c}\displaystyle
	E_{\rm 2t}(x)=E_{\rm i}e^{-jk_0\sin\theta_{\rm i}x}+\sum_{n=-\infty}^{\infty}A_ne^{-jk_{{\rm r}x}x},\\ \displaystyle
	H_{\rm 2t}(x)=E_{\rm i}\cos\theta_{\rm i}e^{-jk_0\sin\theta_{\rm i}x}/Z_0 -\sum_{n=-\infty}^{\infty}A_n\frac{k_{{\rm r}z}}{\omega\mu_0}e^{-jk_{{\rm r}x}x}.
    \end{array}
\end{equation}
By enforcing the boundary condition, $E_{\rm 2t}(x)/H_{\rm 2t}(x)=Z_{\rm s}(x)$, we determine the amplitudes of all  the Floquet harmonics using  the method reported in~\cite{hwang2012periodic} (a MATLAB code can be downloaded from~Ref. \cite{suppl}).
Figure~\ref{fig:fig2a} shows the determined complex magnitudes  of   harmonics  for  illumination at an angle $\theta=-45^\circ$ [hereafter, as an example, we consider a metasurface designed for $\theta_{\rm i}=45^\circ$, i.e., with the  periodicity $D_x=\lambda_0/(2\sin 45^\circ)$]. Three scenarios are illustrated, corresponding to   three different surface impedance profiles~[Eq. (\ref{eq:surface impedance})], which were calculated assuming $R_1=0.5$, $R_1=0.75$, and $R_1=0.99$.
As is seen from Fig.~\ref{fig:fig2a}, for all three scenarios,  the magnitudes of propagating harmonics $n=0$ (specular refection) and $n=1$ (retroreflection with amplitude $R_2$) are zero, meaning that  all of the incident energy is dissipated in the lossy metasurface. The reason of full absorption is excitation of an infinite set of evanescent harmonics. Note that  all the excited evanescent modes  are of negative order ($n\leq-1$) and propagate along the $-x$ direction. This effect highlights the role of   evanescent fields as a mechanism of phase matching between the incident wave and the metasurface. 
Interestingly, the less absorption we require for the illumination at  $+45^\circ$ (the higher $R_1$), the stronger evanescent harmonics are  excited for the illumination at  $-45^\circ$, ensuring full absorption for that illumination direction. 
We can see that the contrast ratio of absorption, defined as $\ln [(1-|R_2|^2)/(1-|R_1|^2)]$, can be arbitrarily adjusted from zero to infinity by manipulating the amplitudes of evanescent harmonics.
In all the three cases for different $R_1$, the  absorptance curve is extraordinarily asymmetric with respect to the incident angle, as shown in Fig.~\ref{fig:fig2b}. The limiting case of $R_1=0.99$ is of special interest where the absorptance curve resembles the Heaviside step function. 
At the critical angle $\theta_{\rm c} =\arcsin(\lambda_{\rm i}/D_x-1) \approx +25^\circ$, where the  mode $n=-1$ becomes propagating or evanescent, the metasurface abruptly switches its response  from an absorber to  a reflector.
Additionally, the frequency-domain response of the surface is discussed in Ref.~\cite{suppl}.


\begin{figure*}[tb!]
	\centering
	\subfigure[]{\includegraphics[width=0.3\linewidth]{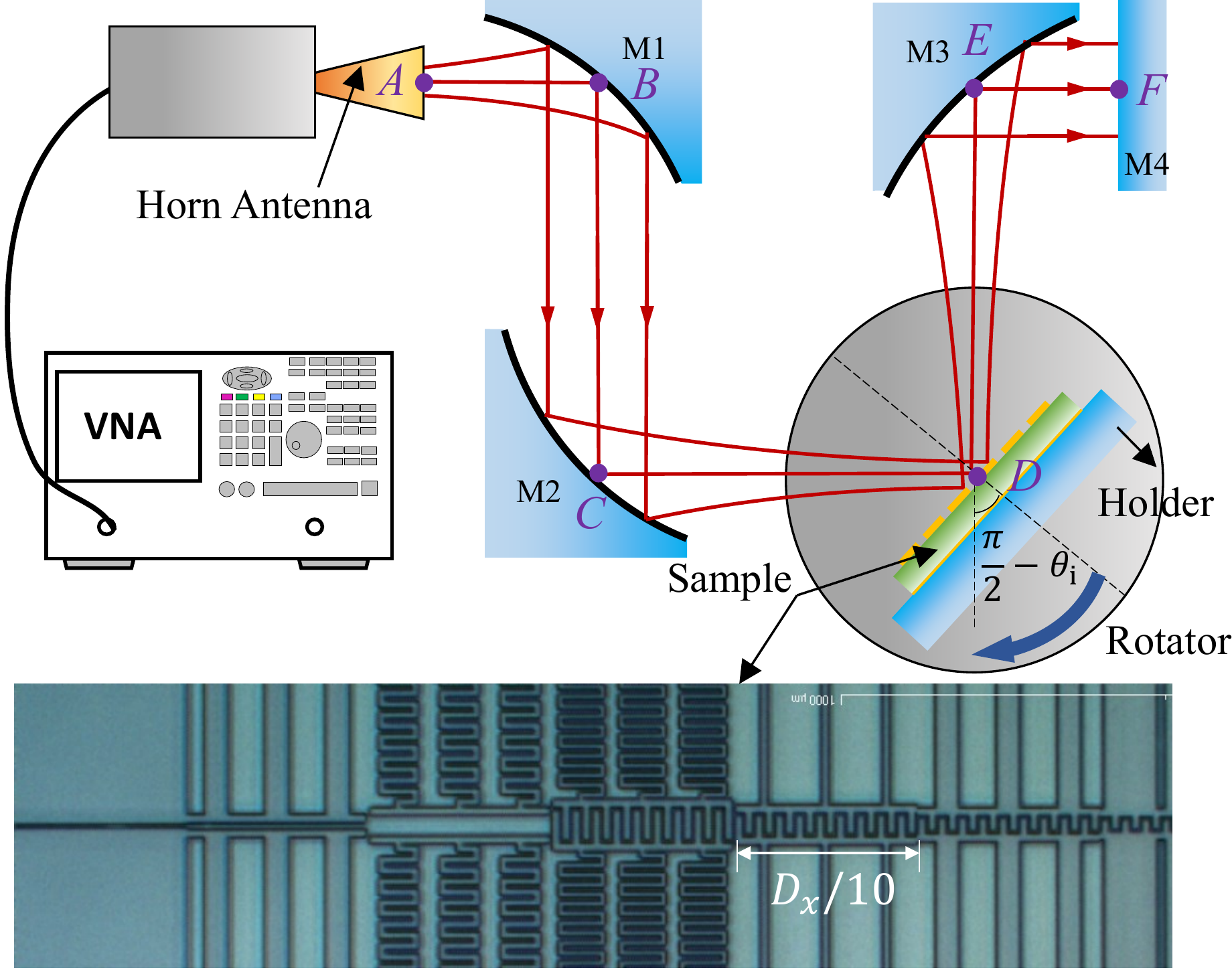} \label{fig:fig3a}}
	\subfigure[]{\includegraphics[width=0.3\linewidth]{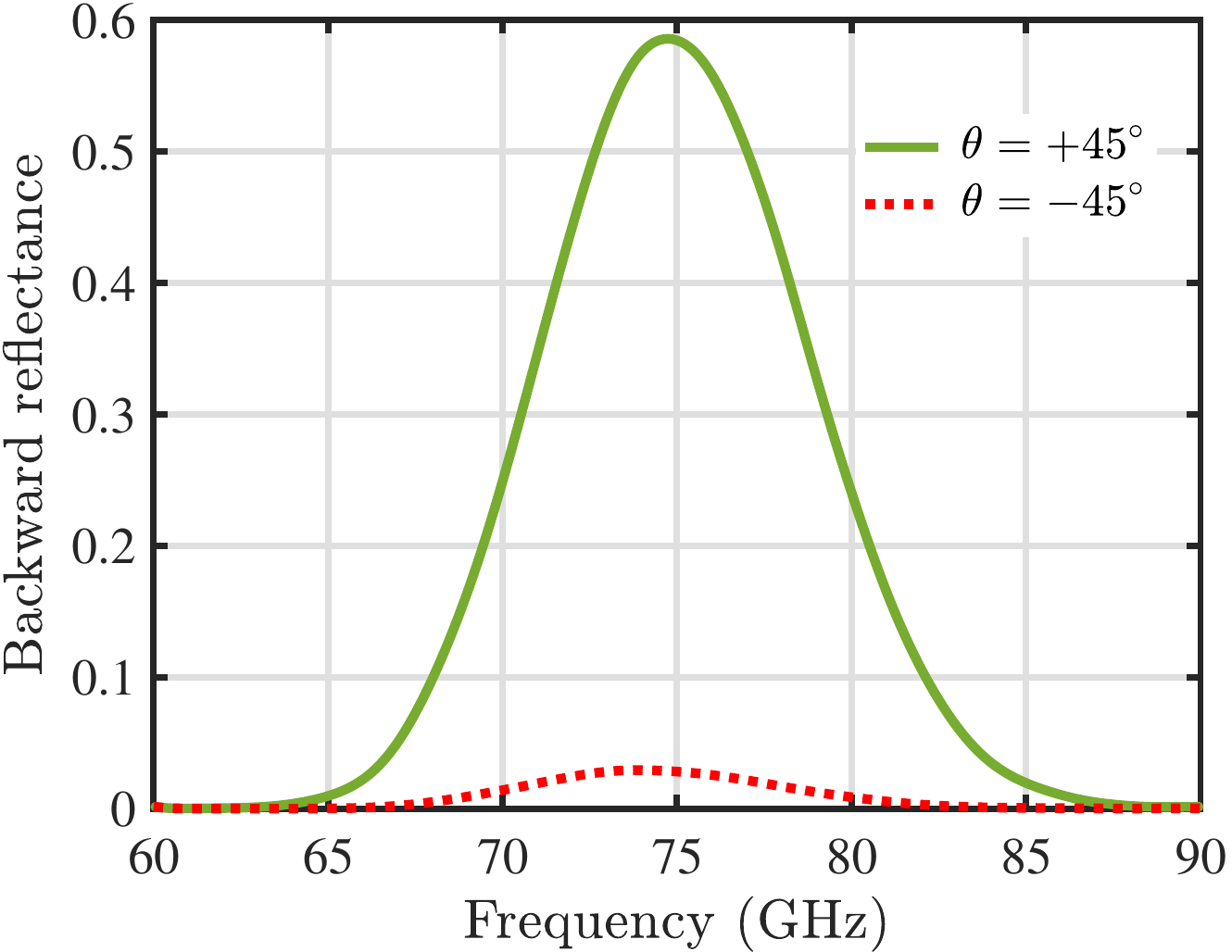}\label{fig:fig3b}}
	\subfigure[]{\includegraphics[width=0.3\linewidth]{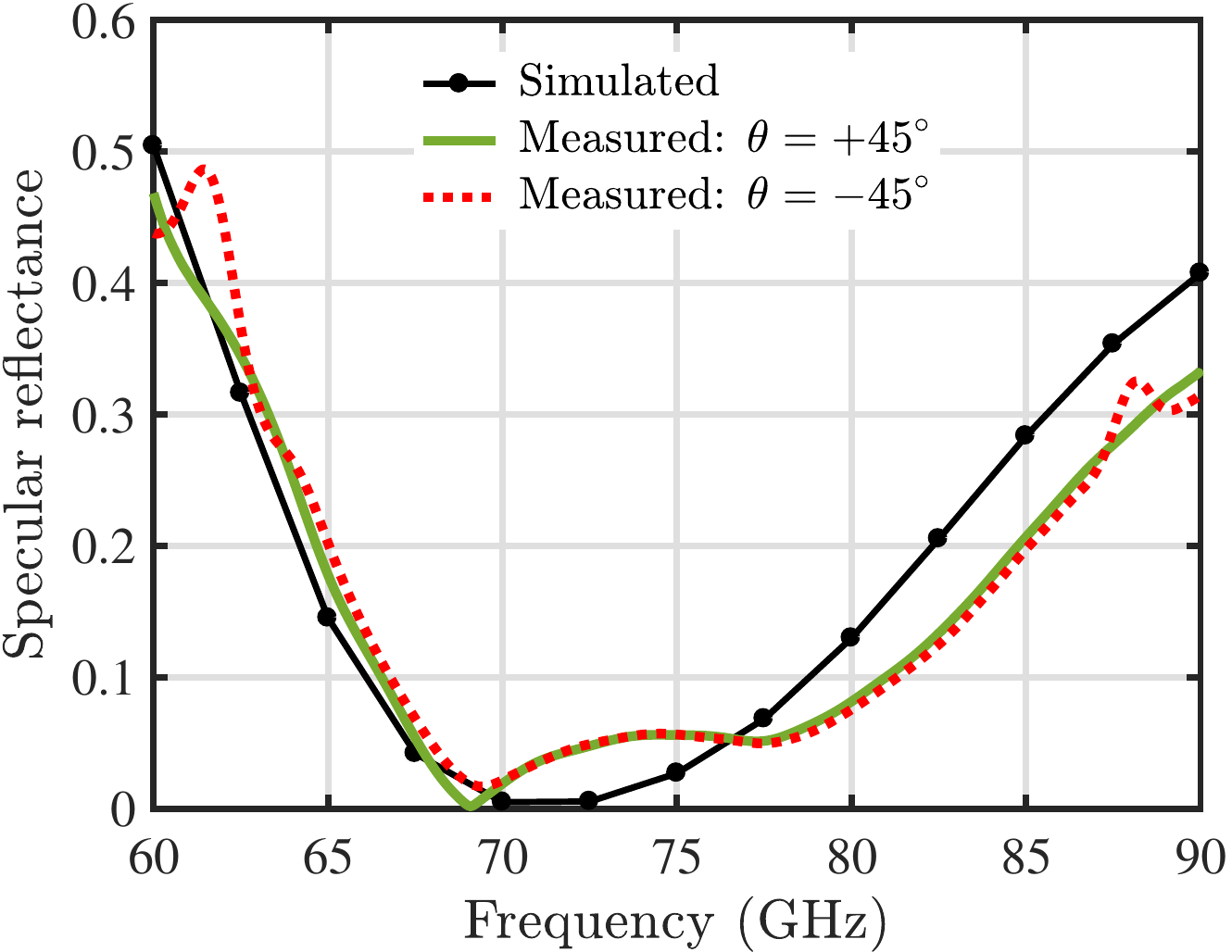} \label{fig:fig3c}}
	\caption{ (a) Top: schematic   of the experimental setup. Bottom: microscope photograph  of the fabricated sample. Measured  (b) backward reflectance (retroreflection squared) and  (c) specular reflectance when the metasurface is illuminated at opposite angles.}
\end{figure*}

Before the implementation of the surface impedance~[Eq.~(\ref{eq:surface impedance})], it is important to consider the influence of the impedance discretization on the metasurface performance. Figure~\ref{fig:fig2c} demonstrates how absorptance for illuminations at angles $+45^\circ$ and $-45^\circ$ depends on the number of sub-cells~$M$ in each period (surface impedance is modeled by a step function with $M$ steps over a period). The data in the figure corresponds to the scenario when $R_1=0.75$; the discretization data for scenarios with other values of $R_1$ are plotted in Ref.~\cite{suppl}. Thus,  the asymmetry of absorptance strongly  depends   on the number of subelements and for higher asymmetry (higher $R_1$) a large number of small sub-cells is required. This fact shows that angle-asymmetric absorption is not possible to achieve using conventional gratings and ``metagratings''. 
In our implementation, we synthesize the surface impedance with $R_1=0.75$ using  $M=10$ sub-elements over a period. 
Such discretization ensures the proper excitation of evanescent waves over the metasurface  (see detailed analysis of evanescent modes versus $M$ in Ref.~\cite{suppl}). 

Next, we realize the required surface impedance at 75~GHz with a metallic pattern supported by a metal-backed dielectric slab [with the relative permittivity $\epsilon_{\rm r}=(4.3-j0.015)$  and thickness $h=215$~$\mu$m].
Based on the transmission-line theory, the required  sheet (or grid) impedance $Z_{\rm g}$ of the pattern can be expressed as~\cite{wang2018systematic} $Z_{\rm g} (x) = [{Z_{\rm s}^{-1}(x)}+jZ_{\rm d}^{-1}\tan^{-1}(k_{\rm n}h)]^{-1}$,
where $Z_{\rm d}=Z_0/\sqrt{\epsilon_{\rm r}-\sin^2\theta_{\rm i}}$ and $k_{\rm n}=k_{\rm 0}\sqrt{\epsilon_{\rm r}-\sin^2\theta_{\rm i}}$ are the wave impedance and the normal component of the wave number in the dielectric. 
The calculated required grid impedance is shown in Fig.~\ref{fig:fig2d}. Using full-wave simulations~\cite{hfss}, we analyze the reflected field from the ideal grid impedance discretized into 10~subcells on the metal-backed dielectric substrate [see Fig.~\ref{fig:fig2e}]. The field distribution confirms that with $M=10$ subcells per period the metasurface possesses desired response: Retroreflection with the field amplitude of $0.74E_{\rm i}$ for illumination at    $+45^\circ$ and nearly full absorption (99.5\%) for illumination at $-45^\circ$. In the latter case,  strong evanescent fields are triggered in the vicinity of the metasurface in accordance with the theory, which induces total absorption with the presence of metal losses. It should be mentioned that the unilateral excitations of strong evanescent waves can have also other applications, for example, for creating nonreciprocal devices based on substrates with nonlinear properties.  
  

Next, we implement the complex grid impedances as 10~subcells.
Here, we utilize a thin aluminum film with thickness of 25~nm and measured grid resistivity of $2$~$\Omega$/sq~\cite{wang2017accurate}. By structuring the homogeneous film, we can create the required grid resistance and reactance on a surface as shown in Fig.~\ref{fig:fig2f}. The required resistance is realized by tailoring the width and length of the metallic strips. For the polarization of the incident wave shown in Fig.~\ref{fig:fig2f}, the smaller width and the longer length result in higher grid resistance.
Since such structures inherently contain some inductive characteristics (which increase when width decreases), we introduce a capacitive gap as another degree of freedom to control the grid reactance. Similarly, the capacitance of the gap can be increased by narrowing or meandering the gap, without affecting the resistive part. In this way, we can independently engineer the grid resistance and reactance. Following this method, all the subcells were optimized with geometrical dimensions specified in Ref.~\cite{suppl}. Figure~\ref{fig:fig2d} (see circle points) shows that the realized grid impedances of the 10 subcells closely follow the required theoretical curve. 
Next, we combine together   the sub-cells into one unit cell   shown in Fig.~\ref{fig:fig2f}. Without further optimization, the simulated results show  retro-reflection  $R_1=0.76$, which is very close to the design target value of  $0.75$, and the reflection   for the opposite illumination is $R_2=0.16$ (97.4\% absorption). The specular reflectance for both illuminations is 2.7\%. 


 Figure~\ref{fig:fig3a} depicts the schematic of the experiment setup (see more details in Ref.~\cite{suppl}). The sample is fabricated using photolithography and lift-off process. Figure~\ref{fig:fig3b} shows the measured backward reflectance spectra (retroreflection squared)  of the gradient metasurface   illuminated at  $+45^\circ$ and $-45^\circ$. The results are in excellent agreement with the theory. At the operating frequency of 75~GHz, the measured backward reflectance for $\theta=+45^\circ$ reaches 58.5\% (versus theoretical $R_1^2=0.75^2\approx 0.56$), while for $\theta=-45^\circ$ it is equal to  2.8\% (versus theoretical $R_2^2=0$). Figure~\ref{fig:fig3c} shows the specular reflectance  for both illuminations. Due to reciprocity of the metasurface,  the two measured curves are practically  identical.  At 75~GHz, only 5.6\% of the incident power is specularly reflected   (versus simulated 2.7\%). Thus, the absorptance calculated from the measured data for illuminations  at $+45^\circ$  and $-45^\circ$ is approximately 36\% and  92\%, respectively.
Due to the fabrication limitations and material dispersion, one cannot infinitely scale down the structure dimensions to higher frequencies towards the visible range.  However, the developed theory is rather general and applicable to the whole electromagnetic spectrum. It is worth  mentioning that the presented structuring method is not a unique solution for the implementation of required impedance profiles. In optics, it is always possible to simplify the configurations of subelements or use high-permittivity dielectric  cells to satisfy the required impedance boundary condition. 

To summarize, we have demonstrated that evanescent waves excited in subwavelength gratings can be exploited to achieve extreme electromagnetic effects in planar structures. As a particular example of an effect that is not possible in conventional phase-gradient metasurfaces or diffraction gratings, we have demonstrated angular-asymmetric absorption and reflection.  The concept of wave control via evanescent harmonics engineering can be scaled to other frequencies and applied to wave processes of  different nature. In addition to its  theoretical significance, it  can have  multiple practical outcomes. For example, by replacing the dielectric substrate of the metasurface by a nonlinear material, one can design various  compact nonreciprocal devices. Since strong evanescent waves are excited only for one of the opposite illuminations, the effective permittivity of the nonlinear substrate will be different for these two cases. As a result, the metasurface response will not obey the Lorentz reciprocity theorem.  Importantly, the amplitude of the  waves impinging on the metasurface  do not have to be large and can be the same for the opposite illuminations. Other important applications of the proposed angular-asymmetric structures include unidirectional emission, multifunctional gratings, systems for one-side detection, sensing and radar cross section control.

This work was supported by the Academy of Finland (Projects 288145, 287894, and 309421). The authors thank Jingwei Zhou (Aalto University) for technical discussions and Tapio M${\rm \ddot{a}}$kel${\rm \ddot{a}}$ (VTT) for his help in the fabrication.

\bibliography{references}
\end{document}